# A unipolar head gradient for high-field MRI without encoding ambiguity

Markus Weiger[1], Johan Overweg, Franciszek Hennel[1],

Emily Louise Baadsvik[1], Samuel Bianchi[1], Oskar Björkqvist[1], Roger Luechinger[1], Jens Metzger[2],

Eric Michael[1], Thomas Schmid[1], Lauro Singenberger[1],

Urs Sturzenegger[3], Erik Oskam[3], Gerrit Vissers[4], Jos Koonen[4],

Wout Schuth[5], Jeroen Koeleman[5], Martino Borgo[5]

Klaas Paul Pruessmann[1]

[1]*Institute for Biomedical Engineering, ETH Zurich and University of Zurich, Zurich, Switzerland*

[2]*Institute for Energy and Process Engineering, ETH Zurich, Zurich, Switzerland*

[3]*Philips AG, Zurich, Switzerland*

[4]*Philips Healthcare, Best, The Netherlands*

[5]*Futura Composites BV, Heerhugowaard, The Netherlands*

| | |
|---|---|
| Corresponding author: | Dr. Markus Weiger |
| Address: | Institute for Biomedical Engineering |
| | ETH Zurich and University of Zurich |
| | Gloriastrasse 35 |
| | CH-8092 Zurich |
| | Switzerland |
| Phone: | +41 44 632 45 44 |
| Fax: | +41 44 632 11 93 |
| E-mail address: | weiger@biomed.ee.ethz.ch |
| | |
| Short running head: | A head gradient without encoding ambiguity |
| Word count (body text): | 2900 |
| Figures: | 8 |
| Tables: | 2 |
| References: | 77 |

**ABSTRACT**

**Purpose:** MRI gradients with a conventional, bipolar design generally face a trade-off between performance, encoding ambiguity, and circumventing the latter by means of RF selectivity. This problem is particularly limiting in cutting-edge brain imaging performed at field strengths ≥ 7T and using high-performance head gradients.

**Methods:** To address this issue, the present work proposes to fundamentally eliminate the encoding ambiguity in head gradients by using a unipolar z-gradient design that takes advantage of the signal-free range on one side of the imaging volume. This concept is demonstrated by implementation of a unipolar head gradient for operation at 7T.

**Results:** Imaging in phantoms and in vivo demonstrates elimination of backfolding due to encoding ambiguity. At the same time, the unipolar design achieves efficiency on par with conventional bipolar design, resulting in high amplitude and slew-rate performance.

**Conclusion:** The prospect of gradient systems based on a unipolar design holds promise for all advanced neuroimaging that demands high gradient performance. It will make the greatest difference at 7T and beyond where the absence of ambiguity removes a key concern and constraint in terms of RF behaviour and instrumentation.

**Key words:** bipolar, backfolding, aliasing, nerve stimulation, brain MRI, 7T

**INTRODUCTION**

Z-axis gradient coils traditionally follow the principle of a Maxwell pair, comprising two sections that generate field of the same spatial structure but opposite polarity (1,2). The two field lobes superimpose to form a bipolar field with a zero in the iso-centre and a surrounding linear range. Outside this range the field reaches maximum excursions, beyond which it gradually drops back to zero, causing ambiguity of gradient encoding. To prevent backfolding in images, the unambiguous range must be made sufficiently large (1). However, unambiguous range comes at substantial expense in terms of gradient performance and amplifier requirements as well as peripheral nerve stimulation (PNS) (2,3). Therefore, the gradient range required to prevent backfolding is traditionally contained by limiting the spatial coverage of RF transmission and detection (4).

While long-established and successful for clinical whole-body systems, this approach to gradient ambiguity is less favourable for advanced brain imaging, which increasingly relies on field strengths of 7T and beyond (5), where RF fields (for $^1$H) are less contained. Higher Larmor frequency renders RF wavelength in tissue significantly smaller than body dimensions, entailing substantially more complex, load-dependent RF behaviour (6-8). In particular, it comes with increasing RF propagation within the body and along the patient bore (9,10). RF complexity and propagation can be tackled by array transmission and detection as well as optimization of coil geometry and shielding (11-23). However, containing RF fields and sensitivity competes with primary RF design objectives such as SNR, coverage, uniformity, and specific absorption rate (SAR) (24), and thus comes at their expense.

At the same time, neuroimaging calls for ever-higher gradient performance, which is increasingly implemented through head-only gradients. Reaching beyond the performance available from whole-body systems was the lead motive when head gradients were conceived (25-31). Continuous improvements led to a recent generation of head gradient systems with greatly enhanced specifications, duty cycle, and practical usability (32-41), paving the way for a range of frontier applications (42-49). This potential was more recently considered also for high-field systems (50-54) and use with a focus on diffusion imaging (55). However, the performance gain with head gradients intrinsically comes at the expense of unambiguous range.

The desire to boost both $B_0$ and gradient performance for neuroimaging thus faces the dilemma that it doubly undercuts the traditional strategy of addressing gradient ambiguity. Related backfolding has in fact been reported for imaging with head gradients at 7T (56-58) (Figure 1). As discussed above, the backfolding can be addressed by added gradient and RF requirements. However, this approach comes at an expense in terms of overall imaging performance, which will likely escalate with field strength.

To address this issue, the present work proposes an alternative approach to gradient ambiguity. It takes advantage of the fact that, when imaging the head, ambiguity is of concern only in the neck and body and thus only on one side of the imaging volume. Gradient encoding without ambiguity can therefore be performed with a unipolar rather than bipolar z-gradient field, effectively reducing the Maxwell pair to one of its halves. The unipolar concept is employed here to implement a head gradient for imaging at 7T. Successful suppression of any backfolding is verified by phantom and in vivo experiments.

**METHODS**

**Unipolar gradient design**

Figure 2 illustrates the described ambiguity issue and the proposed solution for a head gradient. The ambiguity of a conventional, bipolar design of a z-gradient is eliminated by generating a unipolar field instead. In this way, backfolding from the trunk and arms into the imaging band is excluded, irrespective of the characteristics of the transmit and receive RF coils involved. For image encoding, the unipolar field exhibits an approximately linear range like the bipolar template. However, while the latter is zero at the centre of the linear range, the unipolar field has a finite strength at the centre. Within the linear range, this situation is equivalent to off-centre use of a conventional bipolar gradient.

**Simulations**

To demonstrate the ambiguity effect and support the design procedure, simulations of head MRI were performed using full 3D signal encoding and image reconstruction based on the properties of an existing RF coil and hypothetical head gradients.

The human body as a signal source with realistic RF weighting was generated experimentally by imaging a volunteer in a 7T scanner using the standard body gradient and an RF head coil (see below for experimental details). To enable mapping of signals stemming from body locations beyond the unambiguous z-range of the body gradient, data acquisition was performed at six table positions at a distance of 50 mm by starting from the default position and moving the patient bed (together with the RF head coil) further into the bore. The reconstructed imaging volumes were merged subsequently. For scan parameters see Table 1.

Imaging simulations were performed on a 3D grid with isotropic spacing of 10 mm. After slab excitation along the z-dimension, 3D Cartesian encoding was applied for an isotropic FOV of 300 mm using discrete Fourier transform incorporating calculated gradient fields. Images were reconstructed by fast Fourier transform, followed by gradient non-linearity correction (59). To enable independent observation of signals stemming from inside and outside the linearity volume (LV) of the gradient, the signal source was split up accordingly, and two separate simulations were performed.

**System specifications and environment**

A head gradient system for advanced neuroimaging at 7T was developed based on unipolar design of the z-gradient along with bipolar design of the x- and y-gradient fields. The design combines a conical shoulder section for improved patient access (29,40) with a cylindrical main section specified at 390 mm free bore diameter to accommodate typical head RF equipment. In the LV of $220\times220\times200$ mm$^3$, centred at the magnet's iso-centre, field deviation from linearity was limited to 20% (in terms of spatial derivative), corresponding to the equivalent limit on resolution loss.

Targeting high gradient amplitude and rapid switching, the system was designed to operate with switchable dual-mode gradient amplifiers (Copley 787, Analogic Corporation, Peabody, Massachusetts, USA). In parallel mode, (650 V, 720 A), it delivers gradient strength of 200 mT/m, available simultaneously with 280 mT/m/ms slew rate. In the serial mode (1300 V, 360 A), it achieves slew rate of 560 mT/m/ms simultaneous with amplitude of 100 mT/m.

The gradient system was integrated with the magnet, RF, cooling, patient-table and console subsystems of a 7T whole-body imaging setup (Philips Achieva) (Figure 3A). The patient table was modified to fit into the gradient bore and to carry RF coils, interfaces and cables, with the latter exiting the magnet bore at the service side.

**Electromagnetic design and manufacturing**

Optimisation of a z-gradient with unipolar design was performed with the same boundary-element method as otherwise employed for bipolar fields (60,61). With the given specifications, this led to the field shown in Figure 2C. The field offset in the LV centre is 0.124 mT per 1 mT/m gradient strength. For the x- and y-dimensions a conventional bipolar design was employed. To contain current density, a system of double layers of conductors was chosen and optimised to achieve force, torque, and impedance balancing, taking into account the magnet-specific $B_0$ distribution and active shielding. Cooling is based on a combination of hollow and solid conductors. The layer scheme is illustrated in Figures 3B-C. A $B_0$ compensation coil and a full third-order shim set were included in the space between the gradient and shield coils. The gradient was manufactured using dedicated tooling for conductor shaping but otherwise standard state-of-the-art production technology. It is shown in Figures 3D-E.

**Electromagnetic characterisation**

Before installation, the net resistance and inductance were measured for each axis. Furthermore, the field along the z-axis for x = 0 and y = 0 was mapped with low AC current using a fluxgate magnetometer (Mag03 IE 70-3000, Bartington Instruments Ltd, Oxford, UK). After installation, the efficiencies were determined through imaging a phantom of known dimensions. Vector maps of the gradient fields were calculated and stored for analysis and use in image reconstruction, particularly for non-linearity correction.

**PNS**

The uncommon field characteristics of the unipolar design render a PNS investigation particularly important. Thresholds for PNS were determined experimentally in three volunteers (62). Trains of 50 bipolar trapezoidal gradient shapes with plateau duration 0.1 ms were repeated four times at intervals of 200 ms. They were applied on a single axis at a time with the shortest rise time possible for maximum strength and increasing amplitude. A 1.5 s break was inserted between stimuli to give the volunteer the opportunity to report PNS. The procedure was performed in both serial and parallel mode of the gradient amplifier.

**Imaging**

To address the field offset associated with the unipolar design, the software of the scanner was modified to apply the equivalent off-centre of 124 mm in the z-dimension, corresponding to modulation and demodulation of RF signals (63). Apart from the imaging acquisitions, this concerned scan start-up procedures, image reconstruction, and image processing. In particular, iterative shimming with frequency adaptation per increment to keep the position of the excited shim volume, was modified accordingly.

Imaging was performed using an RF head coil (Nova Medical, Wilmington, Massachusetts, USA), including a transmit-receive quadrature birdcage and either a 16- or a 32-channel receive array. The scan parameters are listed in Table 1. Image reconstruction included correction for gradient non-linearity (59). This correction was verified on data obtained from a cylindrical 3D-printed grid phantom (170×170×160 mm$^3$) filled with doped water. Elimination of gradient ambiguity by unipolar design was verified by imaging in phantoms and in vivo.

All experiments in humans were performed according to applicable ethics approval.

**RESULTS**

**Simulations**

Figure 4A shows how at 7T, signals from the neck, trunk, and arms are picked up by the head coil. Although the amplitudes are relatively small, such signals can accumulate to significant image intensity due to only minor encoding through gradient fields. This is demonstrated in Figure 4B with simulated imaging using a bipolar head gradient. Signals stemming from outside the LV are aliased into the LV centre with amplitudes comparable to the targeted signal from the brain. Using instead a gradient with unipolar design, no such artefacts occur.

**Electromagnetic characterisation**

The measured electromagnetic properties of the manufactured gradient are listed in Table 2. Gradient non-linearities and efficiencies match the targeted specifications and are comparable for the bipolar x- and y-gradients and the unipolar z-gradient. The measured field plotted in Figure 5 shows the targeted unipolar field characteristics, and the strong agreement with the calculated field confirms accurate manufacturing.

**PNS**

No PNS was reported by any of the volunteers even at maximum gradient performance in both parallel and serial mode.

**Imaging**

Imaging with the unipolar gradient in presence of the field offset was successfully performed with all protocols, including start-up and reconstruction procedures. The results in Figure 6 obtained with the grid phantom show that distortions due to the gradient non-linearities are successfully corrected by corresponding image processing. The phantom experiment in Figure 7 demonstrates the main feature of the unipolar gradient design, namely that no backfolding occurs from the neck and trunk region into the LV. A slight contamination is observed, which is not due to backfolding but arises from spreading of intense signal through the side lobes of the point-spread-function (PSF). In accordance, imaging of the head in Figure 8 shows no sign of backfolding into the LV. In this case, no signal spreading through the PSF is observed, indicating that the signal distribution of the in vivo situation is rather more benign than in the phantom setup.

**DISCUSSION**

The presented simulations and imaging experiments prove that with the unipolar gradient design no backfolding occurs that impairs the targeted depiction fidelity. Hence, the proposed concept fundamentally solves the ambiguity issue inherent to conventional bipolar gradient fields. The problem thus solved was reported in Refs. (56-58) showing prominent backfolding in brain imaging at 7T. In other literature on 7T imaging with head gradients, the issue has not been apparent (52,64), reflecting viable trade-offs at this field strength. In the design of bipolar z-gradients, the unambiguous range can be increased at the expense of gradient performance and/or amplifier and cooling requirements. In the design and use of RF arrays, sensitivity to backfolding can be addressed at the levels of coil geometry and electronics as well as pulse design and image reconstruction. This added objective, in turn, reduces the degrees of freedom available for optimizing sensitivity, coverage, parallel imaging performance, and SAR. Unipolar design of z-gradients overcomes this conflict and disentangles gradient and RF design such that each can focus on its individual challenges for best performance and economics. This advantage grows fundamentally more significant as field strength increases beyond 7T (65-71), further reducing RF wavelength, increasing the tendency of RF fields to propagate, and complicating RF design in general.

Notably, gradient ambiguity is also linked to the specifications of main magnets in that static off-resonance outside the uniform volume of a magnet helps prevent backfolding (56). This coupling of gradient and magnet design is equally overcome by a unipolar z-gradient.

Unlike conventional gradients, the unipolar design exhibits a finite field strength at the LV centre. This property has previously occurred in other contexts and for different purposes. It has been a side effect of shifting the LV towards the patient end (33,72) and is intrinsic to fringe-field imaging (73) as well as single-sided MRI devices (74).

As demonstrated in this work, the field offset can readily be handled by corresponding modulation and demodulation of RF transmit and receive signals, respectively. For excitation and acquisition under constant gradients, the offset only shifts carrier frequencies but does not affect RF bandwidths. With non-Cartesian (e.g., radial) or curvilinear (e.g., spiral) k-space trajectories, the carrier frequency varies from shot to shot or during readouts. In these cases, depending on the flexibility of the spectrometer used, the field offset can be met by time-varying centre frequency or suitably expanded bandwidth. Besides actual imaging procedures, the field offset must also be accounted for in calibration, scan preparation, and correction procedures that involve or relate to the z-gradient.

Unipolar design entails near-zero field strength in the torso along with a field maximum approximately twice as large as with bipolar design. This increase in maximum field drives up energy density, which is proportional to field strength squared. Greater net field energy generally comes at the expense of coil efficiency. This disadvantage is countered by an efficiency gain due to the fact that the intrinsically more distal unipolar coil is largely unconstrained by the shoulders and can be almost fully implemented on a small radius. In the design phase of this work, these two effects were found to be approximately balanced, resulting in roughly equal efficiency for unipolar and bipolar designs of otherwise equal specifications. This observation is supported by comparison with the bipolar design reported in Ref. (52,75) with z-gradient efficiency of $\eta = 0.16$ mT/m/A and inductance of $L = 315$ μH, which translate into an efficiency factor (3,76,77) of $\varepsilon = 0.0813$ mT/m$^4$/A while $\varepsilon = 0.193$ mT/m$^4$/A has been achieved here with unipolar design (see Table 2). These values are not directly comparable because the gradients have different inner diameter (44 cm vs. 39 cm). Correction by diameter to the power of five (3) puts the advantage of the present unipolar design at approximately 30%, at comparable linearity (75). This indicates that unipolar design is clearly competitive in terms of efficiency.

Unipolar design for head imaging has important consequences also with regard to PNS and the risk of myocardial stimulation. Importantly, with vanishing field at the location of the heart, unipolar design fundamentally overcomes any risk of myocardial stimulation, which is a significant advantage over conventional z-gradients. For the same reason, the unipolar z-gradient is very benign with regard to PNS in the neck and shoulders. Concern about PNS due to higher maximum field on the opposite side is offset by the fact that the largest field values occur only well outside the body. Nevertheless, increased PNS is to be expected at the top of the head. In the reported tests, fully exploiting the switching capability of the present gradient implementation did not cause any PNS. Favourable PNS properties, as described in the initial abstract on this work (51), have recently motivated a unipolar implementation of an ultrasonic gradient coil (49).

**CONCLUSION**

Unipolar design has been confirmed to solve the ambiguity problem of head-only z-gradient coils. It does so by moving from the three field ramps of a Maxwell pair to exposing the body to only the single field ramp required for gradient functionality. In addition to ambiguity, this approach eliminates the consideration of myocardial stimulation as a potential risk. Despite an increase in field maximum, a unipolar gradient is favourable in that it generates most field energy outside the body where coil geometry is not constrained by the subject. As a result, net coil efficiency at least on par with bipolar design has been achieved. Elimination of gradient ambiguity frees RF considerations from the need to support the suppression of backfolding, lifting a complex constraint from RF optimization. This advantage grows more relevant as the complexity of RF behaviour and the demand for large channel counts increase with field strength. Therefore, unipolar gradients hold particular promise for high-performance neuroimaging at 7T and beyond.

Weiger et al., A head gradient without encoding ambiguity

**TABLES**

**Table 1**

Imaging parameters. Dimensions of FOV and resolution are given in the order x, y, z. Abbreviations: GE, gradient echo; GS, gradient-spoiled; T1w, T1-weighted.

| Protocol | Object | Figure | Sequence | FOV [mm] | Resolution [mm] | TR [ms] | TE [ms] | Flip angle [°] | Receive coil | Scan time [m:s] |
|---|---|---|---|---|---|---|---|---|---|---|
| Signal source | Human body | 4 | 3D GS-GE | 400×500×530 | $2.5^3$ | 5 | 2 | 6 | 32-ch. array | 5:36 |
| Non-lin. correction | Grid phantom | 6 | 3D GS-GE | $200^3$ | $1.0^3$ | 10 | 1.6 | 15 | Birdcage | 5:15 |
| Ambiguity | Oil bottles | 7 | 3D GS-GE | $300^2 \times 512$ | $1.0^3$ | 6 | 1.6 | 15 | Birdcage | 11:47 |
| Ambiguity | Human head | 8 | 3D T1w-GE | 189×252×300 | 0.8×1.0×0.8 | 6 | 2 | 15 | Birdcage | 5:58 |

**Table 2**

Electromagnetic properties of the unipolar gradient. The field deviation was determined as the maximum of the field error as defined in Ref. (3). The efficiency factor ε was calculated according to Ref. (3) where the factor 2 was omitted as in Refs. (76,77). G = gradient strength, I = current.

| Axis | Gradient deviation [%] | Field deviation [%] | Resistance @ DC [mΩ] | Inductance L @ 1 kHz [μH] | Efficiency $\eta = G/I$ [mT/m/A] | Efficiency factor $\varepsilon = \eta^2/L$ [mT/m$^4$/A] |
|---|---|---|---|---|---|---|
| x | 16.6 | 6.0 | 88 | 541 | 0.279 | 0.144 |
| y | 16.6 | 6.0 | 81 | 483 | 0.279 | 0.161 |
| z | 17.4 | 4.2 | 87 | 400 | 0.278 | 0.193 |

Weiger et al., A head gradient without encoding ambiguity

**FIGURES**

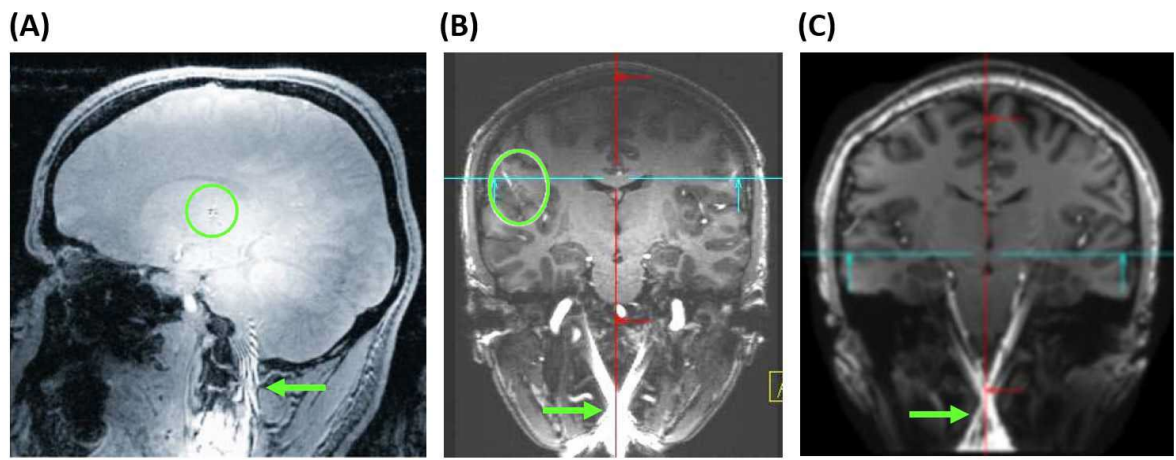

**Figure 1**

Examples of backfolding artefacts at 7T due to ambiguous encoding with a bipolar head gradient. Artifacts occur in the lower head (green arrows) as well as in central brain regions (green circles). Reproduced with permission from A) Ref. (56), B) Ref. (57), and C) Ref. (58).

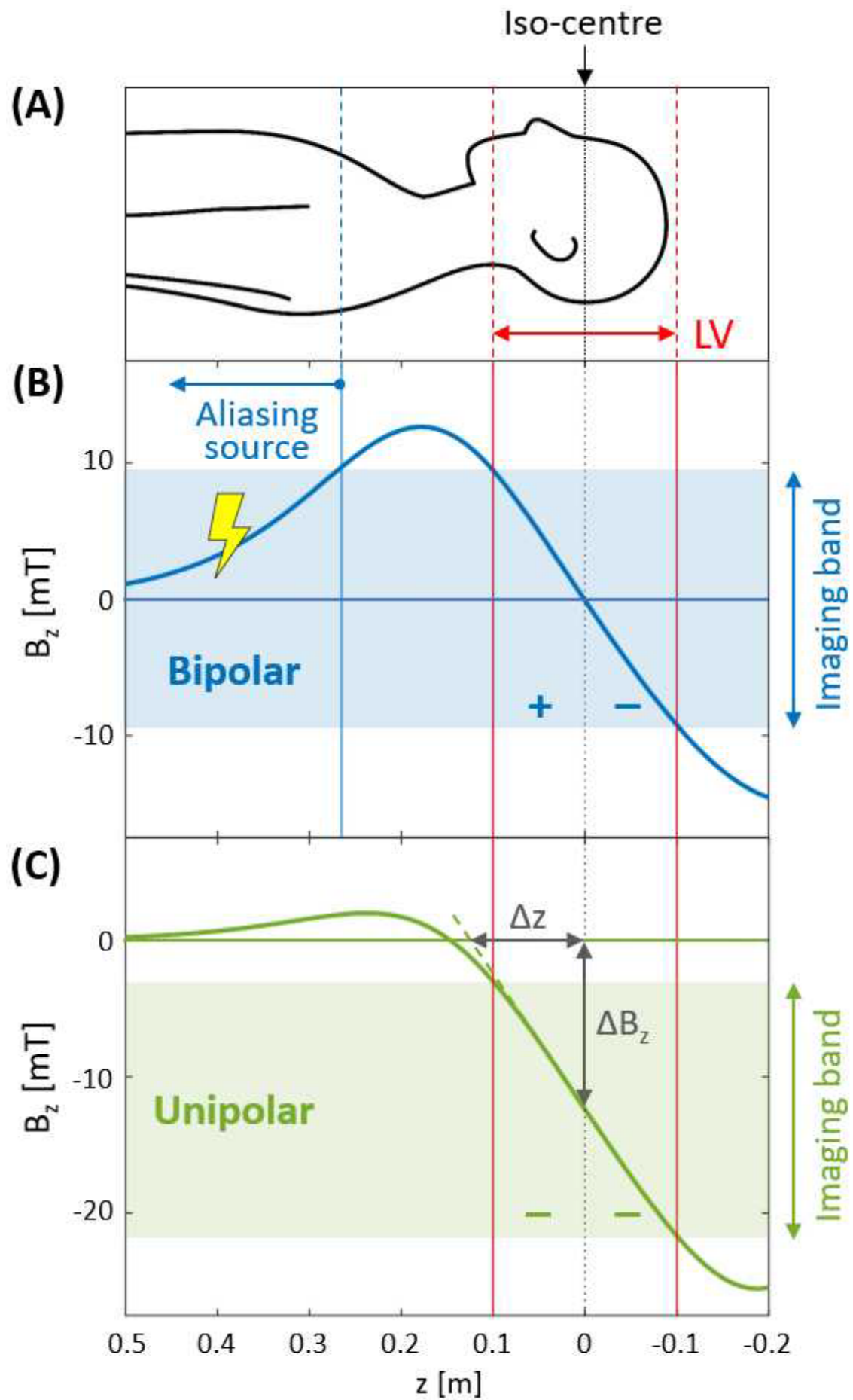

Weiger et al., A head gradient without encoding ambiguity

**Figure 2**

Ambiguity issue and solution for a head gradient. (A) Human body with its brain centred at the magnet's iso-centre at z = 0. The linearity volume (LV) of the gradient covers the imaging target volume. (B) Gradient coils for the z-axis traditionally follow the principle of a Maxwell pair (1). With the associated bipolar field, signals picked up by the RF coil from the "aliasing source" in the body region are encoded undistinguishably from those in the LV, and are hence aliased into the "imaging band". (C) This ambiguity is eliminated in the unipolar design, where the Maxwell pair is effectively reduced to one of its halves. The remaining ambiguity on the other side of the imaging volume is of no concern, as there are no signal sources present. The approach introduces an additional field offset $\Delta B_z$ in the LV, which is proportional to the applied gradient strength and therefore creates a situation equivalent to imaging with an off-centre $\Delta z$. As indicated, $\Delta z$ is given by the zero-crossing of the ideal gradient field (dashed green line). The $B_z$ field values given in the plots are based on a gradient strength of 100 mT/m.

Weiger et al., A head gradient without encoding ambiguity

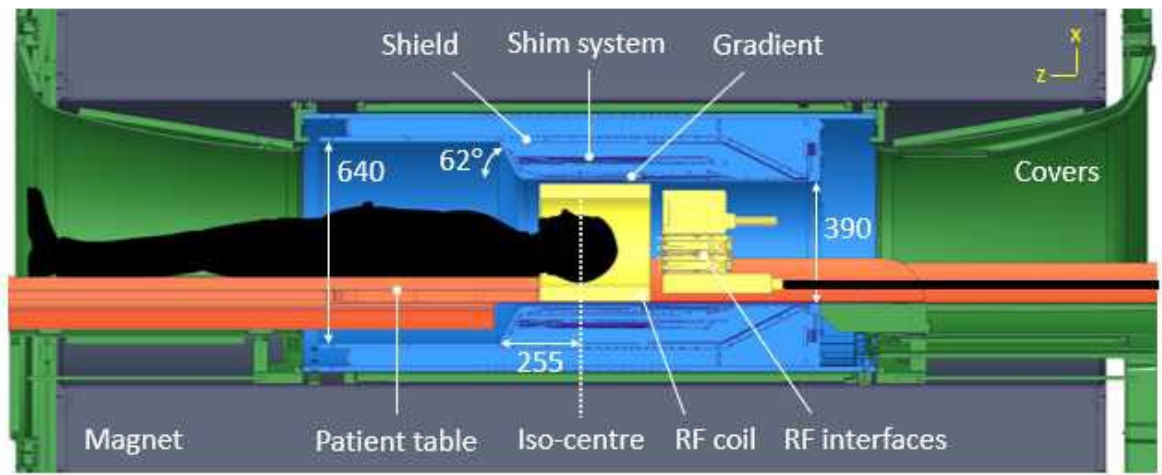

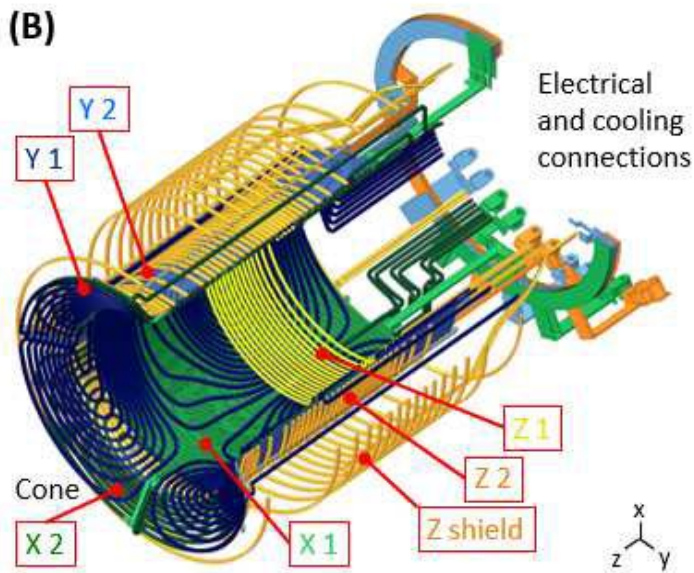

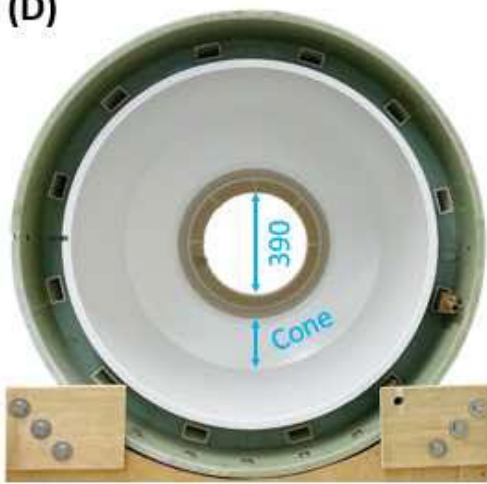
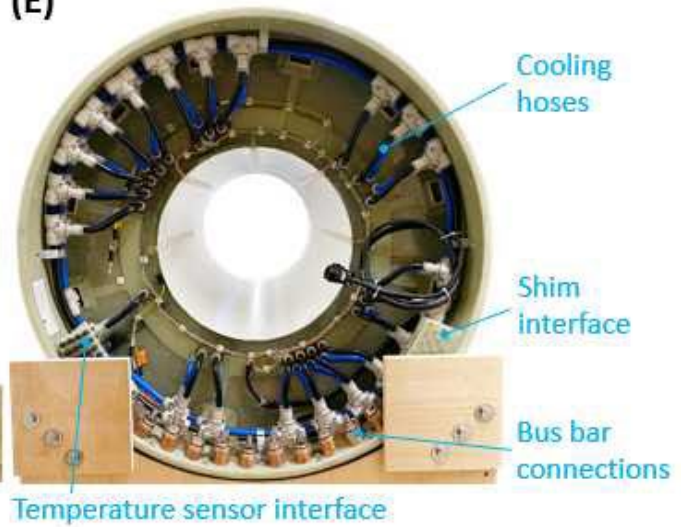

**Figure 3**

Implementation of a unipolar head gradient for 7T. (A) Drawing of the gradient (blue) inside the 7T magnet (grey), showing the main dimensions (mm) and the locations of gradient, shim, and shield coils, as well as the conical widening in the shoulder region to allow patient access to the iso-centre. The patient table is modified to fit into the bore and carry RF coil, interfaces, and cables. (B) Conductor layout, showing the two layers of all axes and the z-shield, including electrical and cooling connections at the service side. Shields for x and y are not shown. (C) Layer scheme indicating which of the coils are wound from a hollow conductor for active cooling and which have a conical part. (D) Patient and (E) service end of the manufactured gradient with indications for the inner bore and the conical opening, as well as for interfaces for bus bars, cooling, shim supply, and temperature monitoring using 32 fibre-optic sensors.

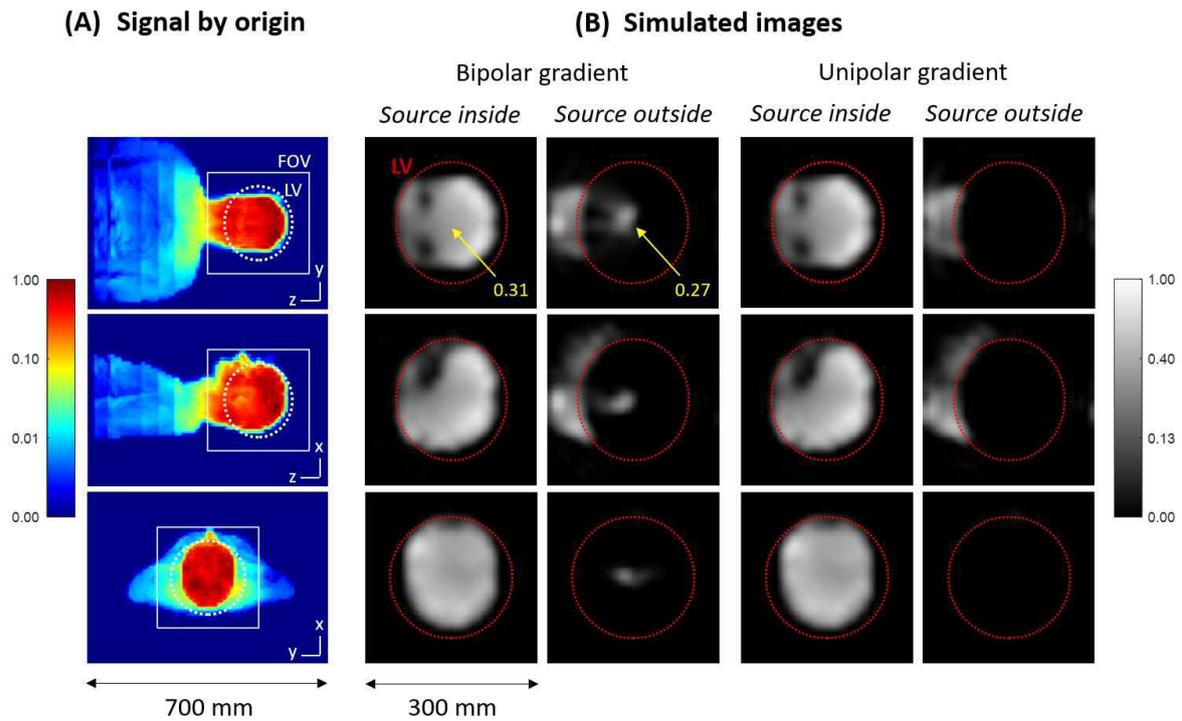

**Figure 4**

Simulations of 3D MR imaging using bipolar and unipolar head z-gradients (see Figure 2). (A) Experimentally determined signal source by origin as used in the simulations, showing signals picked up by the RF head coil at 7T from neck, trunk, and arms. The linearity volume (LV) of the gradient and the imaging FOV used in the simulations are indicated. (B) Simulations performed separately for signals stemming from inside and outside the LV. Orthogonal images are displayed as maximum-intensity projections and with the logarithmic scaling shown by the colour bars. With the bipolar gradient, signals from the body are aliased into the LV centre, with an amplitude comparable to the brain signal. Using instead the unipolar gradient, no such backfolding occurs. Signal sources from the lower head and neck are correctly depicted outside the LV.

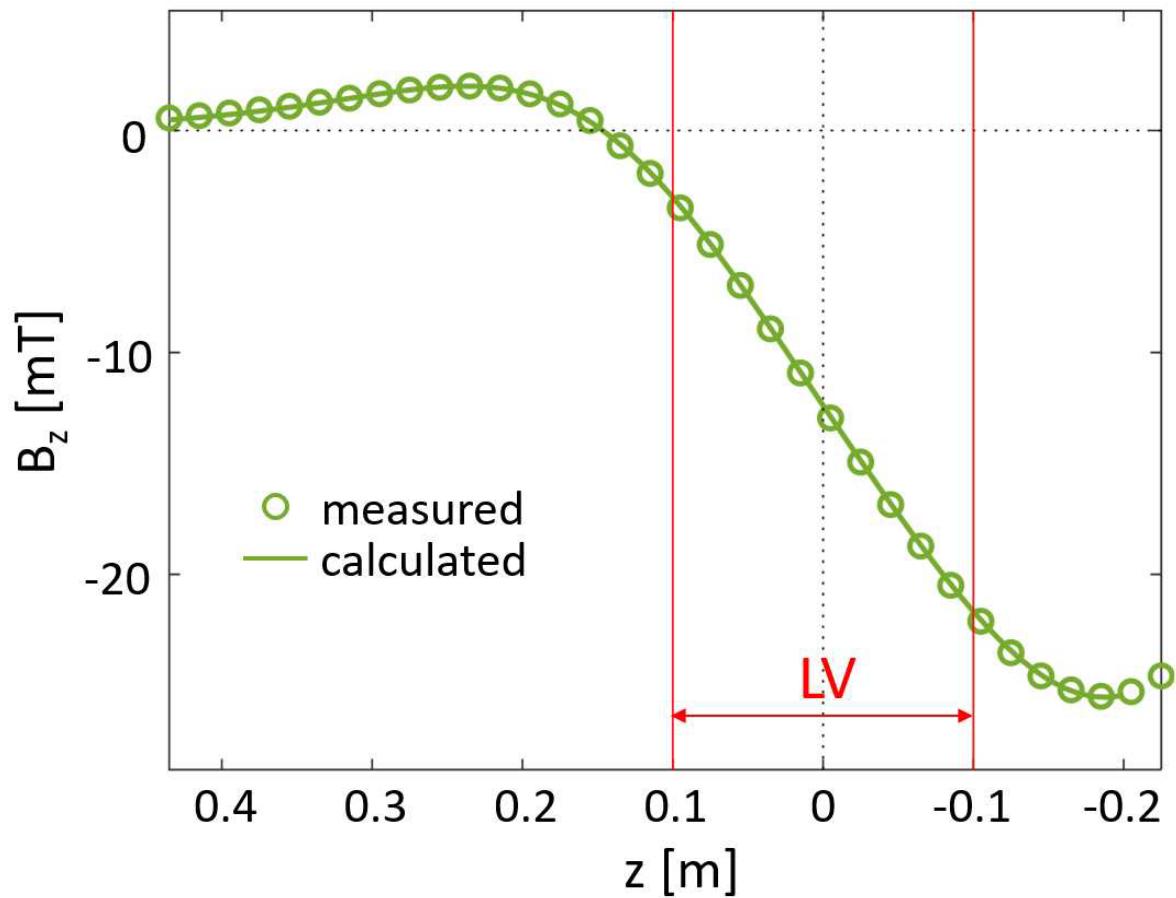

**Figure 5**

Field plot of the manufactured z-gradient along x = 0 and y = 0. After scaling the data measured at low current to a gradient strength of 100 mT/m, it is virtually identical to the calculated unipolar target field.

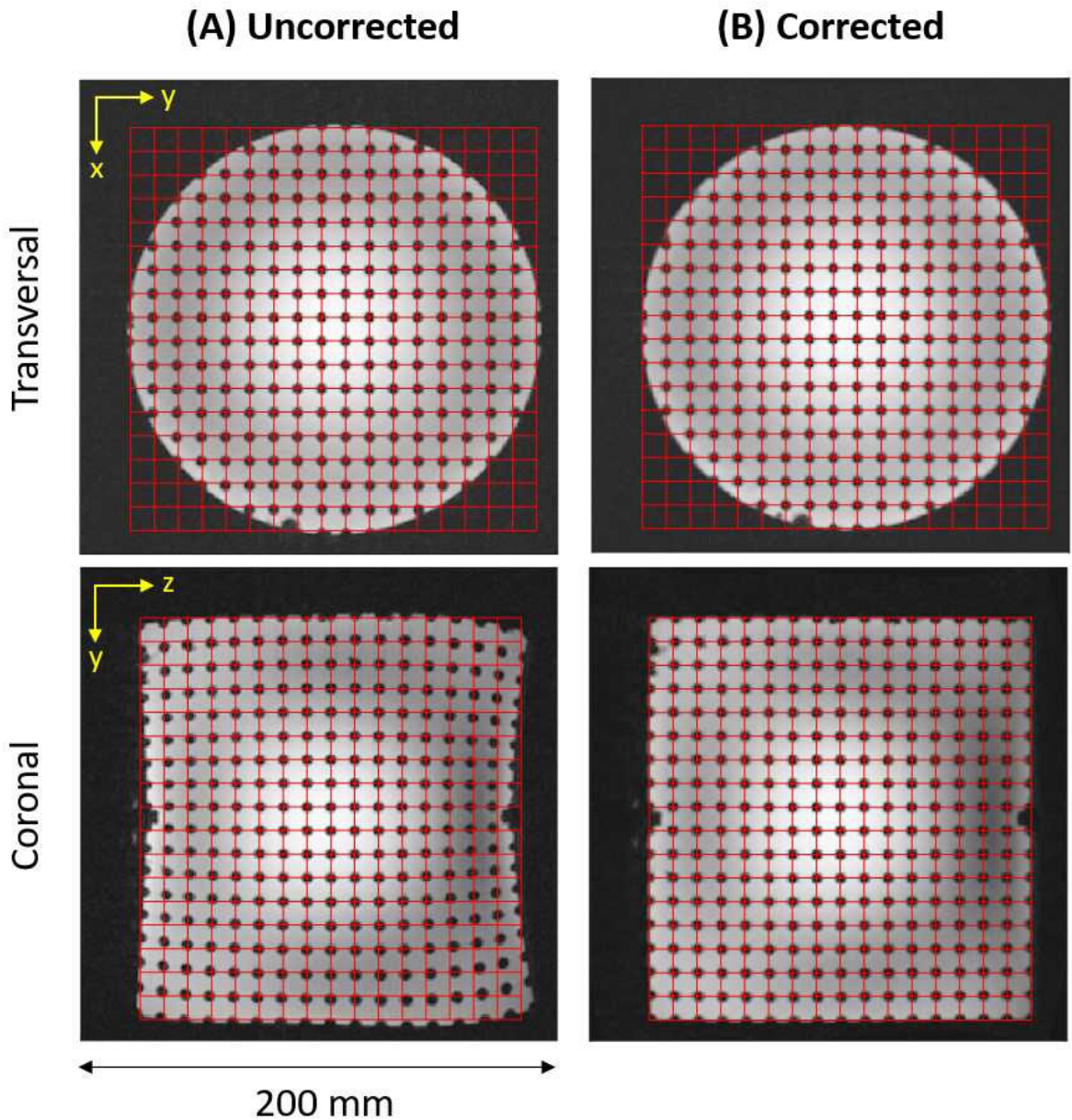

**Figure 6**

Gradient non-linearity correction. Central transverse and coronal slices acquired from a grid phantom are displayed with an overlaid grid (red) to indicate the expected node positions. (A) Uncorrected images show distortions which are most pronounced at larger z-positions. (B) With non-linearity correction based on calculated field maps, the distortions are essentially eliminated. Note that image intensities are displayed with logarithmic scaling to reduce typical $B_1$ non-uniformities at 7T. See Table 1 for imaging parameters.

Weiger et al., A head gradient without encoding ambiguity

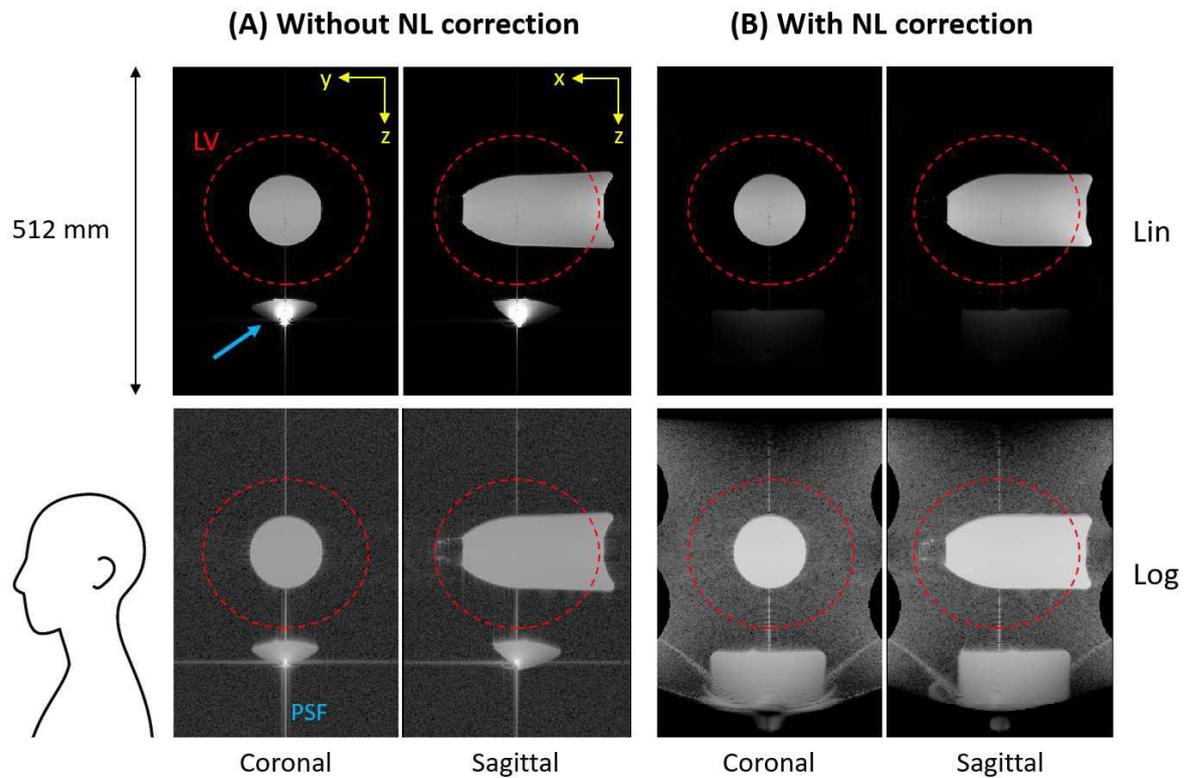

**Figure 7**

Phantom experiment demonstrating the absence of backfolding from the neck and body region into the linearity volume (LV). A 1l oil bottle was placed in the LV centre and a 5l oil bottle was located just outside the RF head coil to mimic signal stemming from neck and trunk. Central coronal and sagittal slices from the 3D data set are displayed with linear and five-fold logarithmic intensity scaling, where the latter serves to enable observation of low-intensity signals. (A) Weakly encoded signal from locations outside the LV is concentrated in a small volume (blue arrow), but does not reach into the LV. With logarithmic scaling it is observed that the main signal spike is spread through the side lobes of the 3D point-spread-function (PSF). (B) By non-linearity (NL) correction, the concentrated signal is largely un-warped and the shape of the bottom part of the larger bottle can be recognised. The lower end of the signal-containing region indicates the limit of the unambiguous range and coincides with the plateau of the field maximum in Figure 2. There are no signs of backfolding into the LV. See Table 1 for imaging parameters.

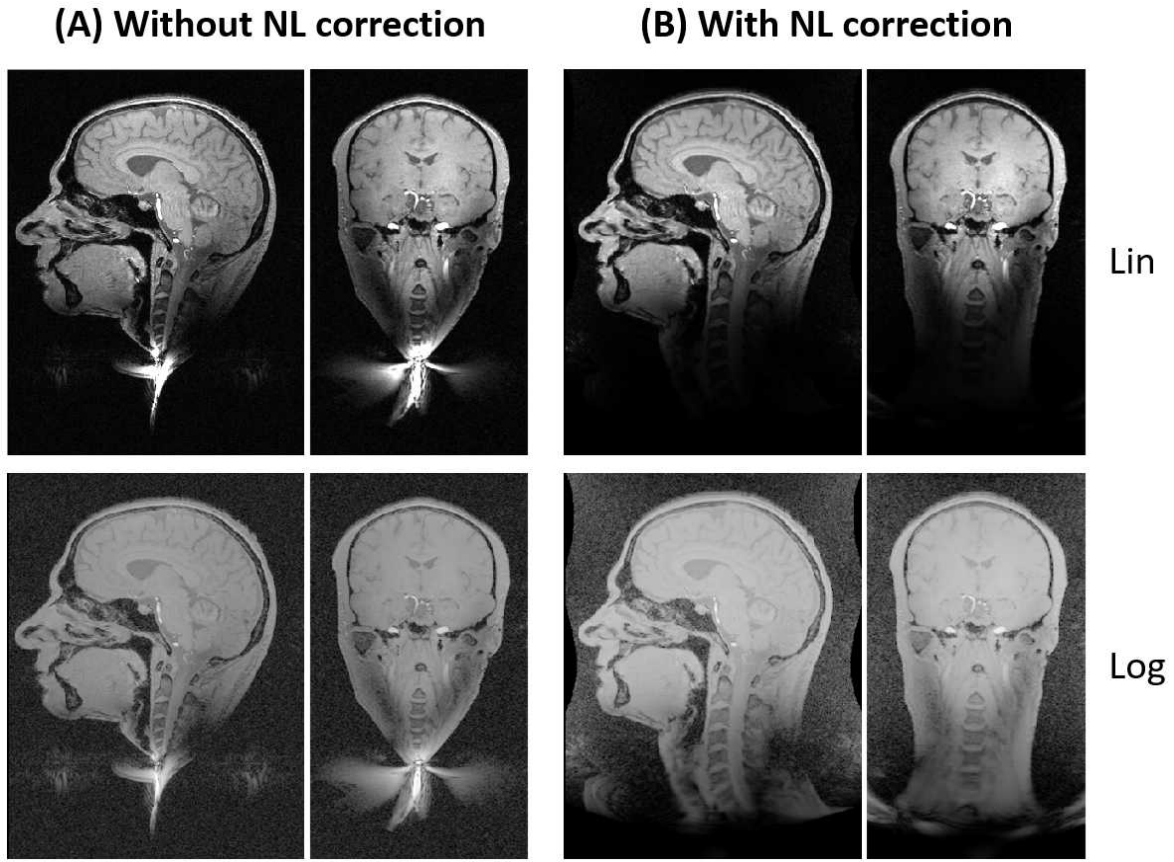

**Figure 8**

In vivo head imaging. Central coronal and sagittal slices of the 3D data set are displayed with both linear and logarithmic intensity scaling. (A) The pattern of concentrated signal below the head reflects the weakened gradient fields beyond the LV. (B) This signal is largely un-warped by the non-linearity (NL) correction. No signs of aliasing into the head from the neck and body region are observed. Moreover, the PSF-induced signal spreading as seen in the phantom experiment in Figure 7 is absent. See Table 1 for imaging parameters.